\documentstyle[preprint,aps]{revtex}
\begin{document}
\renewcommand{\bottomfraction}{1.0}
\renewcommand{\topfraction}{1.0}
\renewcommand{\textfraction}{0.0}
\renewcommand{\floatpagefraction}{0.9}
 
\title
{Metallic screening and correlation effects in superconducting fullerenes}
 
\author{P.E. Lammert}
\address{Groupe de Physico-Chimie Th\'eorique, ESPCI, F-75231 Paris Cedex 05, 
France}
\author{D.S. Rokhsar}
\address
{Department of Physics, University of California, Berkeley, CA 94720}
\author{S. Chakravarty and S. Kivelson}
\address
{Department of Physics, UCLA, Los Angeles, CA 90024}
\author{M.I. Salkola}
\address
{Theoretical Division, Los Alamos National Laboratory, Los Alamos, NM 87545}
 
\date{July 14, 1994}
 
\maketitle
\begin{abstract}
 
We consider the {\em frequency dependent} Coulomb interaction between 
electrons in a molecular metal in the limit in which the conduction 
bandwidth is much less than the plasma frequency, which in turn is 
much less than intramolecular excitation energies.  In particular,
we compute the effective interactions at the Fermi energy
in alkali-doped C$_{60}$, to second order in the screened interactions.
The frequency dependence of the screening substantially reduces the 
effects of the long-range part of the Coulomb interaction, leading to 
the possibility of an effective attraction between electrons.
\end{abstract}
\pacs{74.20.-z, 74.70.Wz, 71.27.+a}
 
The alkali- and alkaline-earth-doped fullerenes\cite{review} are
molecular solids, characterized by strong intra\-molecular 
electron-phonon and electron-electron interactions, but weak 
inter\-molecular hybridization.
They can exhibit metallic, magnetic, insulating, or superconducting 
behavior; the $A_3{\rm C}_{60}$ compounds are superconductors 
with transition temperatures as high as 33~K.
 
Electrons at the Fermi energy are strongly coupled to intramolecular
phonons, especially the high frequency tangential modes with energy
$\hbar\omega_c$ $\approx$ 200 meV.
A variety of microscopic calculations\cite{phonons,CGK} yield estimates 
of the BCS effective attraction produced by these intramolecular modes 
of roughly 75 meV-${\rm C}_{60}$.
Since the bands at the Fermi energy are narrow (bandwidth $W \approx$ 
200 meV), the electronic density of states at $E_F$ is correspondingly 
large: $N(E_F)$ $\approx$ 15/eV-spin-${\rm C}_{60}$.  
The dimensionless electron-phonon coupling constant $\lambda \equiv$
$N(E_F) V_{\rm BCS}$ is then of order unity.  
The implications of this coupling are unclear, however, since the 
electronic bandwidth and characteristic phonon energies are comparable, 
preventing the straight\-forward application of the
Born-Oppenheimer/Migdal-Eliashberg approximation.
 
Electron-electron interactions are also important in the doped fullerenes:
the typical intramolecular Coulomb energy is $e^2/\epsilon_\infty R_b$
$\approx 2$ eV, where $R_b=3.5$\AA\ is the fullerene radius.
The importance of Coulomb scattering in the doped fullerenes was 
emphasized by Chakravarty {\it et al.\/},\cite{CGK} who calculated 
the effective interaction at the Fermi energy\cite{LamRok} using an
on-site Hubbard repulsion $U$ to model the screened Coulomb interaction.
Although the bare Hubbard interaction is repulsive,
Chakravarty {\it et al.} found that for moderate $U$ both second order
perturbation theory\cite{CGK} and exact diagonalization studies of small
fullerene analogs\cite{White} exhibit an effective interaction at the
Fermi energy which is anomalously small
or even negative (indicating an effective attraction).
The possibility of an effective attraction suggests an electronic
pairing mechanism due to intramolecular correlation effects.
Much of the phenemenology of fullerenes can be understood in the context
of such correlation effects.\cite{CGK,SCK} 
 
How do longer-range Coulomb interactions affect these considerations?
Goff and Phillips\cite{GofPhi} have shown, also using second order
perturbation theory, that moderate-range ({\it e.g.}, several bond-lengths)
bare interactions lead to effective interactions at the Fermi energy that
are substantially more repulsive.
They argue that an electronic mechanism is thereby ruled out.
Similar arguments have been made by Auerbach and Murthy.\cite{AueMur}
Such arguments misrepresent the complexity of the electron-electron 
interactions in the fullerenes, since such a strong effective repulsion 
would overwhelm the phonon mechanism as well! (There can only be a limited 
renormalization of the electronic repulsion since the phonon frequencies
are comparable to the conduction bandwidth.\cite{comment})
 
While these calculations raise the important issue of long-range
Coulomb effects, the extended Hubbard models of refs. \onlinecite{GofPhi} 
and \onlinecite{AueMur} omit the crucial effects of metallic screening.
In this note we consider a simple, physical model for this {\em frequency 
dependent} screening, and study its effect on the energetics of pairing.  
Our model relies on the separation of energy scales corresponding
to intermolecular hopping ($W$), metallic screening ($\hbar\omega_p$),
and intramolecular hopping ($W_\pi$), as shown schematically in fig. 
\ref{energy-scales} and discussed in detail below.
Although correlation effects cannot be computed exactly, they can be 
studied in a controlled manner using second-order perturbation theory. 
This permits a comparison with the perturbative calculations of refs.
\onlinecite{CGK}, \onlinecite{GofPhi}, and \onlinecite{AueMur}.

We find that a long-ranged but frequency dependent screened interaction
strikingly reproduces the results obtained from the Hubbard model.\cite{CGK}
Frequency dependence is a crucial ingredient in this somewhat surprising 
result, which can be understood heuristically as follows:
The direct repulsive interaction of electrons at the Fermi energy (see 
fig. \ref{diagrams}a) occurs at a frequency below the plasma frequency, and
is therefore heavily screened.  The higher-order corrections (which can be
attractive in second order, fig. \ref{diagrams}b-e), however, involve
excitations above the plasma energy, and are therefore unscreened.  
The weakening of the first order repulsion relative to the 
second order attraction explains the physics of our results.
Similar considerations may also apply to cluster compounds
known as Chevrels,\cite{Chevrel} although these superconductors
are not as strongly molecular as the doped fullerenes.
 
 
{\bf Frequency dependent screening.}
In alkali-doped C$_{60}$, the bands at the Fermi energy are derived from
a set of three symmetry-related molecular orbitals.
Due to their narrow bandwidth, electronic hopping between molecules is
sluggish compared with the screening response of the surrounding metal:
the bandwidth, $W\approx 0.2$eV, is an order of magnitude
smaller than the Drude plasma energy, $\hbar \omega_p \approx 1-1.5$eV.
Interactions within the conduction bands are therefore well screened.
The intra-molecular virtual fluctuations to the rest of the $\pi$-complex,
spread over $W_{\pi}\approx 15$eV,
however, occur too rapidly to be screened by the surrounding metal. The
corresponding Coulomb matrix elements are therefore largely {\em un}screened.
(Intra-molecular plasmon-like excitations at 6 eV of the carbon $2p\pi$
orbitals will be explicitly taken into account below.
Screening due to carbon $2p\sigma$ orbitals can be absorbed into a frequency 
independent dielectric constant because the characteristic energy for such 
excitations is larger than $W_{\pi}$.)
 
The separation of energy scales $W_\pi > \hbar \omega_p > W$ (see fig. 
\ref{energy-scales}) suggests
a simple model for the frequency-dependent screened Coulomb interaction.
We treat the intra-band repulsion at the Fermi energy as fully
screened, since the bandwidth $W$ is considerably smaller than the
plasma energy $\hbar\omega_p$.
Scattering matrix elements to states away from the Fermi energy, however,
cannot be screened by the surrounding metal, since the plasma frequency is
significantly smaller than typical intra-molecular excitation energies.
 
 
We therefore  approximate the screened Coulomb interaction by
its high and low frequency parts:
\begin{eqnarray}
V(r, \omega) =
\cases{V_{\rm lo}(r) \  {\rm if} \  \omega < \omega_p; \cr
       V_{\rm hi}(r) \  {\rm if} \  \omega > \omega_p, \cr}
\label{model-V}
\end{eqnarray}
where $r$ is the distance between a pair of $\pi$ electrons on a 
given fullerene molecule.
 
The {\it high frequency} part of the effective Coulomb interaction is
simply the unscreened Coulomb interaction:
\begin{equation}
V_{\rm hi}(r) = {{e^2} \over {\epsilon r}}.
\label{hi-freq}
\end{equation}
The dielectric constant $\epsilon$ $\approx$ 2 is determined by
the polarizeability of degrees of freedom which are omitted
from our model, especially the carbon $2p\sigma$ complex.
 
The {\it low frequency} interaction $V_{\rm lo}(r)$ between electrons
on a single molecule can be estimated by considering a molecule of
radius $R_{\rm b}$ at the center of a spherical cavity of radius
$R_{\rm c}$ inside a metal (fig. \ref{cavity}).  
This approximation can be justified diagramatically.
The image charges formed in the metal result in a screened interaction
\begin{equation}
V_{\rm lo}(r) = {e^2 \over \epsilon} \left[ {1 \over r} -
{R_{\rm c} \over
\sqrt{
(R_{\rm c}^2 - R_{\rm b}^2)^2 + R_{\rm c}^2 r^2 }}\right].
\label{lo-freq}
\end{equation}
 
 
Since the distance from the center of one molecule to the nearest carbon
nucleus of a neighboring molecule is 6.5 \AA, and the carbon $2p_z$
orbitals extend roughly 1 \AA\ from the nucleus, we estimate the
effective cavity radius to be $R_{\rm c} \approx$ 5-6 \AA.
While the geometric radius of an individual fullerene molecule is 3.5 \AA,
the $2p\pi$ orbitals are squeezed to the outside of the molecule by
its curvature.  The effective radius $R_{\rm b}$ at which the
electrons move is therefore somewhat larger than the geometric
radius of the molecule, perhaps by as much as half an Angstrom.
 
The {\it on-site} interaction is an atomic property
of the carbon $2p$ orbitals, and must be treated separately.
For a carbon $2p$ orbital, the unscreened Coulomb integral
$V_{\rm hi}(0)$ is roughly 7-10 eV.
Metallic screening of this atomic Coulomb integral should be negligible,
so for simplicity we present here results for $V_{\rm lo}(0)$$=$
$V_{\rm hi}(0)$ $\equiv$$V(0),$
where $V(0)$ corresponds roughly to the ``Hubbard $U$.''
Eqns. (\ref{hi-freq}) and (\ref{lo-freq}), plus the high and low frequency
on-site interactions, complete the specification of our model for
frequency dependence in the screened Coulomb interaction of metallic
fullerenes.
 
{\bf Calculation of pairing interaction.}
To calculate the effective interaction arising from eqns. (1-3), 
we start with a simple tight-binding
model for the $\pi$ molecular orbitals of ${\rm C}_{60}$.
The matrix elements $t$ and $t'$ corresponding to nearest-neighbor 
intra-pentagon (``single bond'') and inter-pentagon (``double bond'')
hopping are $2.0$ and $2.6$ eV, respectively.
 
The three partially occupied molecular orbitals in the alkali-doped
fullerenes transform as the t$_{1u}$ representation of the icosahedral
group.
The pair state favored by both the intramolecular phonon and the
electronic pairing mechanisms is the unique spin and orbital
singlet\cite{LamRok}
 \begin{equation}
| A_{g}, S=0\rangle \equiv {1 \over \sqrt{3}}
[ |xx\rangle + |yy\rangle  + |zz\rangle ] ,
\label{singlet}
\end{equation}
where the filled sea is implicitly included and $x$, $y$, and $z$
label the three orthogonal states of $t_{1u}$ symmetry.
 
We calculate the effective singlet pair interaction energy in this
state,\cite{LamRok} which corresponds to the negative of the pair
binding energy of ref. \onlinecite{CGK}.
This and other renormalized interactions can in turn be added to a
model for intramolecular hopping and phonons to describe the
low-energy behavior of the metal and superconductor.\cite{LamRok}
 
Explicitly, we compute the Feynman diagrams shown in fig. \ref{diagrams}.
The double dashed lines represent the screened Coulomb interaction of
eqns. (1-3);
the incoming and outgoing thin lines are propagators for the singlet
pair state eq. (\ref{singlet}).  We sum over all intermediate states
with at least one particle or hole in orbitals away from the Fermi energy.  
Such propagators are represented
by thick lines in fig. \ref{diagrams}.  There are over 60,000 distinct 
terms in this sum.
In this manner, all virtual particle-hole fluctuations involving
the higher-lying states are included to second-order in the interaction.
Our calculation includes (a) the first-order scattering
at the Fermi energy and (b) the usual second-order ladder, as well as 
(c) the crossed ladder, (d) screening from an intra-molecular particle-hole 
bubble ({\it i.e.,} the 6 eV $2p\pi$ intra-molecular plasmon), and 
(e) the leading vertex correction.
Due to the explicit frequency
dependence of our model interaction, one cannot easily sum even the
ladder diagrams, as can be done when frequency dependence is neglected.
 
%
 
Fig. \ref{results} shows the second-order effective interaction
calculated using the screened Coulomb interaction discussed above.
The results are plotted {\it versus} the on-site interaction $V(0).$
Results for cavity radii $R_{c}$ = 5 \AA\ and 6 \AA\ are shown.
For comparison, we show the second-order calculations of Chakravarty
{\it et al.} for the Hubbard model and of Goff and Phillips for 
the exponentially screened interaction $V(r) = U e^{-\lambda r}$.  
In these two models the frequency-dependence of screening effects have
been neglected, and interactions are instantaneous.
 
The effects of the long-range part of the screened Coulomb interaction 
are substantially reduced by the properly accounting for their frequency 
dependence, as seen by comparing our results with Goff and Phillips.'
Within second-order perturbation theory, we find that the effective 
interaction changes sign near $V(0)$ of order 7-8 eV, which is close to 
the value obtained for the Hubbard model.   
The effective Coulomb interactions for the Hubbard model and the present 
frequency-dependent model are quite similar, except at small $U$, where 
the present model includes longer-range Coulomb interactions that yield 
a repulsive effective interaction even when the bare on-site interaction 
vanishes.
 
The suppression of the effects of the long-range part of the
Coulomb interaction can be understood schematically as follows.  In the
usual ladder sum, the first- and second-order terms (fig. \ref{diagrams}a,b)
{\em both} involve the screened Coulomb interaction.  In a molecular metal, 
however, second- and higher-order terms are {\em not} screened, since they 
involve virtually excited states above the plasma frequency.  These 
higher-order terms are therefore stronger ({\em i.e.,} more attractive) than 
one might naively expect.  Thus screening reduces the (repulsive) 
first-order term, but not the (attractive) second-order term.

In conclusion, we stress that strong intramolecular correlation effects
are necessarily present in a molecular metal such as C$_{60}$.  
We have shown that a combination of inter\-molecular screening and 
intra\-molecular correlations leads to a dramatic reduction of the 
Coulomb repulsion between electrons.  Such a reduction is necessary 
for a consistent theory of phonon-mediated superconductivity in the 
fullerenes in the presence of strong electron-electron repulsion.
To the extent that second-order perturbation theory is valid, 
frequency-dependent screening may even result in an effective pair
binding arising purely from electronic correlation effects.

\acknowledgements{ 
This work was supported by the NSF under PYI grant DMR-91-57414 (P.E.L. and 
D.S.R), grant DMR-92-20416 (S.C.), and grant DMR-93-12606 (S.K.).
M.I.S. was supported in part by the U.S.~Dept. of Energy.
}

 \begin{figure}
 \caption[]{\small The electronic excitations of metallic fuller\-enes are
 character\-ized by three energy scales:
 the bandwidth $W$ due to inter\-fullerene hopping, the plasma
 frequency $\omega_p$, and the width of the $\pi$ complex $W_\pi$.
 \label{energy-scales}}
 \end{figure}
 
 \begin{figure}
 \caption[]{\small
 We model the low-frequency effects of metallic screening by
 considering a single C$_{60}$ molecule inside a spherical cavity of 
 radius $R_{\rm c}$ in a metal.
 \label{cavity}}
 \end{figure}

\begin{figure}
\caption[]{\small The first-order contribution to the effective
interaction between two electrons in the spin singlet, orbital
singlet channel is represented
by (a) direct scattering by the screened interaction.
There are four second-order terms:
(b) the uncrossed ladder, (c) the crossed ladder,
(d) the vertex correction, and (e) screening due to on-ball
particle-hole pairs.  In the second order diagrams (b-e), at least
one of the thick propagators must lie away from the Fermi energy.
\label{diagrams}}
\end{figure}

\begin{figure}
\caption[]{\small Second order calculations of the effective singlet
interaction at the Fermi energy.
The two solid lines are the results of the present work, corresponding
to cavity radii $R_{\rm c} =$  5 \AA\ and $R_{\rm c} =$ 6 \AA.
The dotted line is the Hubbard model calculation of ref. \onlinecite{CGK}.
The dashed lines are the results of Goff and Phillips.\cite{GofPhi}
The horizontal axis is the on-site Coulomb repulsion, $V(0)$, or equivalently
the Hubbard $U$.
For comparison, the phonon mediated attraction $V_{\rm BCS}$ is roughly
75 meV.
\label{results}}
\end{figure}
 
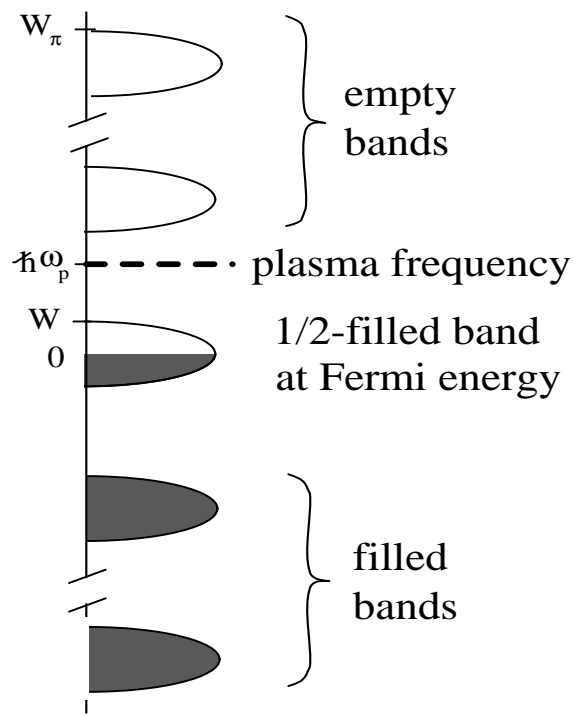
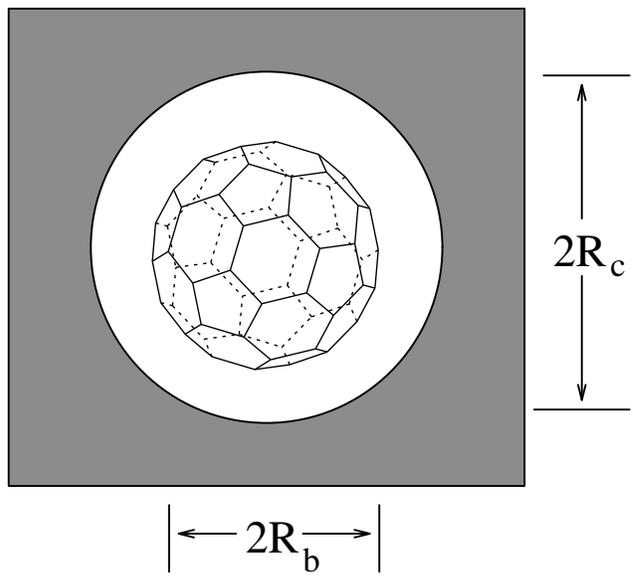
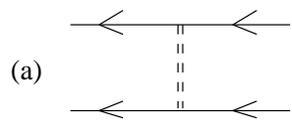
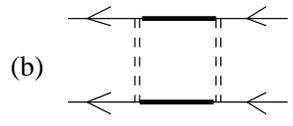
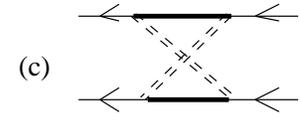
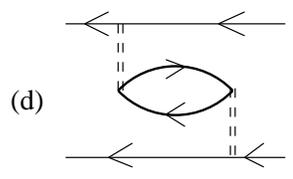
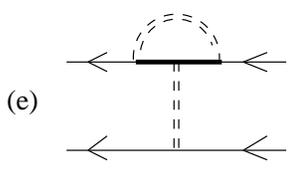
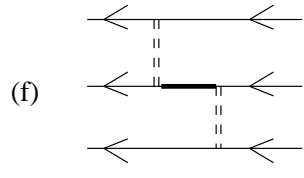
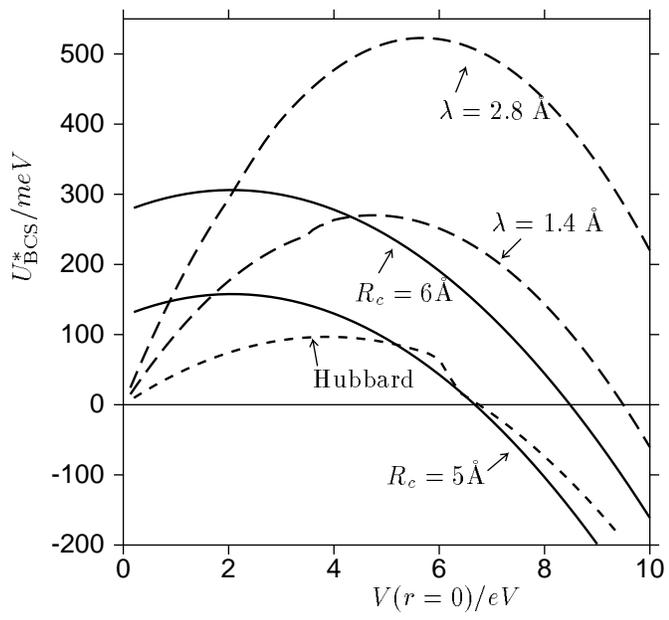
\end{document}